\documentclass[twocolumn,aps,prl,showpieces,superscriptaddress,groupedaddress,floatfix]{revtex4}

\usepackage{comment}
\usepackage{color}
\usepackage{amsmath}
\usepackage{calrsfs}
\usepackage{amssymb}
\usepackage{graphicx}
\usepackage{braket}
\usepackage{esint}
\usepackage{adjustbox}
\usepackage{mathtools}
\usepackage{appendix}
\usepackage{float}
\usepackage[unicode=true,pdfusetitle,
bookmarks=true,bookmarksnumbered=false,bookmarksopen=false,
breaklinks=true,pdfborder={0 0 1},backref=false,colorlinks=true]
{hyperref}
\hypersetup{
	linkcolor=red,urlcolor=blue,citecolor=blue,anchorcolor=blue}
\makeatletter
\@ifundefined{textcolor}{}
{%
	\definecolor{BLACK}{gray}{0}
	\definecolor{WHITE}{gray}{1}
	\definecolor{RED}{rgb}{1,0,0}
	\definecolor{GREEN}{rgb}{0,1,0}
	\definecolor{MAGENTA}{cmyk}{0,1,0,0}
	\definecolor{YELLOW}{cmyk}{0,0,1,0}
}
\usepackage{dcolumn}\usepackage{bm}

\makeatother

\begin{document}
\title{Gain-assisted quantum heat engine based on electromagnetically induced transparency}

\author{Laraib Niaz}  
\affiliation{Quantum Optics Lab, Department of Physics, COMSATS University Islamabad, Pakistan}
\author{You-Lin Chuang}
\affiliation{Physics Division, National Center for Theoretical Sciences, Taipei, 10617, Taiwan}
\author{Fazal Badshah}
\email{fazalbadshah@huat.edu.cn}
\affiliation{School of Electrical and Information Engineering, Hubei University of Automotive Technology, Shiyan, 442000, People's Republic of China}
\author{Rahmatullah}
\email{rahmatktk@comsats.edu.pk}
\affiliation{Quantum Optics Lab, Department of Physics, COMSATS University Islamabad, Pakistan}

\date{\today}
\begin{abstract}
We propose a scheme to realize a gain-assisted quantum heat engine (QHE) based on electromagnetically induced transparency (EIT). The QHE consists of a three-level $\Lambda$-type atomic system that interacts with two thermal reservoirs and a coupling field. The gain without inversion is induced in the system via spontaneously generated coherence (SGC) between two lower levels. The SGC has a significant effect on the system's dynamics, resulting in an enhancement of the emission cross-section and spectral brightness of the QHE.
	
\end{abstract} 

\maketitle
\section{Introduction}  \label{Sec-1}
The QHE has attracted considerable interest due to its potential applications as well as for investigating the quantum limit of classical thermodynamics such as the limit of the Carnot heat engine \cite{kosloff2014quantum}. In 1824, Sadi Carnot proposed the idea of an ideal heat engine that operates by extracting heat from a hot reservoir, converting a portion of it into mechanical work, and then transferring the remaining heat to a cold reservoir. Decades later, with the discovery of the maser, Scovil and Schulz-DuBois introduced the concept of a QHE and demonstrated that its efficiency could be described by the Carnot efficiency  \cite{boukobza2007three}. Both the Carnot and the QHE require at least two reservoirs with differing temperatures,  $T_{C}<T_{H}$, where $T_{C}(T_{H})$ is the temperature of cold (hot) reservoir) and no mechanical work can be done if $T_{C}=T_{H}$.
A novel QHE extracting work even from a single heat bath, i.e., $T_{C}=T_{H} $ was proposed by Scully and his coworkers \cite{scully2003extracting}. They introduced a small amount of coherence in a three-level atomic system and initially prepared the atoms in the superposition of the lower two levels. Since then, there has been a significant advancement in the implementation of QHEs by utilizing various techniques such as the optomechanical effect\cite{zhang2014quantum,gelbwaser2013minimal}, squeezed thermal reservoir \cite{rossnagel2016single}, and superconducting resonators \cite{hardal2017quantum}. Clearly, this is significant progress in the field of thermodynamics as it opens up new avenues for developing efficient heat engines capable of converting heat into mechanical work with improved efficiency and performance.

The efficiency of the QHE in multilevel atomic systems can be increased via quantum coherence \cite{uzdin2015equivalence, dorfman2013photosynthetic} which is the main mechanism behind EIT. EIT refers to a process in which the absorption of a probe field in an optical medium is reduced or suppressed by the presence of a strong control field. As a result, the system appears transparent to the probe field, even though it would normally be optically thick. Recently, a new scheme for QHE was proposed by Harris, based on the concept of EIT \cite{harris2016electromagnetically}. This proposal was soon confirmed experimentally through the demonstration of EIT-based QHE in cooled $^{85}\text{Rb}$ atoms \cite{zou2017quantum}. In this model, a three-level $\Lambda$-type atomic system having a ground, a meta-stable, and an excited state coupled with two thermal reservoirs and one coupling field. A thermal reservoir at temperature $T_{13}$ interacts with the transition from the ground to the excited state. The second thermal reservoir at temperature $T_{23}$ and a coupling field are applied to transition from an excited to a meta-stable state. Initially, a photon is absorbed from reservoir $T_{13}$, and a photon is generated in $T_{23}$ reservoir and the population of the meta-stable state is increased by one unit. Following that, a photon from a coupling field is absorbed and a photon is emitted on the transition from excited to ground level, and the last transition accounts for the mechanical work. The QHE using EIT is also called nontraditional QHE because we can take either $T_{13}$=$T_{23}$ or $T_{13}>T_{23}$ ($T_{13}<T_{23}$). The details working mechanism of EIT-based QHE is given in Ref. \cite{harris2016electromagnetically}.  The performance of nontraditional QHE can also be improved by applying a microwave field between ground and meta-stable state \cite{zhang2019microwave}. A recent investigation of QHE in a Doppler-broadened atomic medium demonstrates that hot atoms can also be used to construct non-traditional QHE \cite{zhang2021doppler}.

Another type of coherence called SGC can emerge in a $\Lambda$-type atomic system when two degenerate lower levels possess non-orthogonal dipole moments  \cite{menon1998effects}. This type of coherence is a result of the quantum mechanical interaction between the two degenerate states, leading to the generation of coherence without any external driving fields \cite{menon1998effects}. SGC has been observed to produce some fascinating outcomes, including the elimination of spectral lines \cite{zhu1996spectral}, enhancement in quantum interference \cite{yannopapas2009plasmon} and gain without inversion \cite{zhou1997quantum}. These results highlight the importance of SGC in the field of quantum optics and its potential applications in areas such as spectroscopy and quantum information processing. Furthermore, the study of SGC can provide new insights into the behavior of quantum systems and inform the design of quantum devices such as quantum heat engines.

In this paper, we suggest a modified version of the nontraditional QHE by incorporating SGC between two lower levels. The emission and absorption cross-sections, which are directly linked to the output field, are found to be significantly influenced by the SGC coefficient. The use of SGC compensates for the losses in the EIT medium and improves the spectral brightness and emission cross-section of the QHE. With the effect of SGC, we demonstrate that the performance of the QHE can be greatly enhanced compared to traditional thermal engines. Furthermore, the role of SGC in improving the power output of the QHE shows its importance in the development of quantum technologies for energy conversion and power generation.

\section{Atomic Model and Equations} \label{Sec-2}
\begin{figure}
    \centering
    \includegraphics[width=8.5cm]{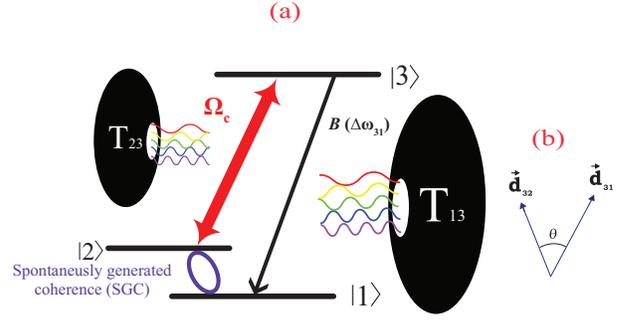}
    \caption{(a)  The schematic of a three-level atomic system. The atomic transition $\ket{1}\rightarrow \ket{3}$($\ket{2}\rightarrow \ket{3}$) interacts with black-body radiations at temperature  $T_{13}$($T_{23}$). The $\ket{2} \rightarrow \ket{3}$ transition is coupled with a monochromatic laser or control field with a Rabi frequency $\Omega_{\text{c}}$. The SGC is generated between $\ket{1} \rightarrow \ket{2}$ transition. As a result of the EIT phenomenon, a strong and spectrally narrow emission is observed in the $z$-direction. This emission has a temperature that is higher than that of any of the reservoirs. (b) The non-orthogonal configuration of dipole moments $\Vec{d}_{31}$ and $\Vec{d}_{32}$.}
    \label{fig:model}
\end{figure}

The schematic of the proposed EIT-based QHE model is shown in Fig. \ref{fig:model}(a). We consider a three-level $\Lambda$-type atomic configuration in which $\ket{1}$ $\rightarrow$ $\ket{2}$ transition is meta-stable whose transition frequency is $\omega_{12}$. A coupling field of frequency $\omega_{\text{c}}$ is applied between transition $\ket{2}\rightarrow \ket{3}$. The Rabi frequency of the coupling field is defined as $\Omega_{\text{c}}=|\Vec{d}_{32}|E_{\text{c}}/2\hbar$, where $E_{\text{c}}$ is the amplitude of the coupling field and $|\Vec{d}_{32}|$ is the dipole matrix element. The incoherent pumping rates ( black-body radiations) $R_{13}$ and $R_{23}$ at temperature $T_{13}$ and $T_{23}$ interact with $\ket{1}$ to $\ket{3}$ and $\ket{2}$ to $\ket{3}$ transitions respectively. The quantum heat cycle can be described as 
$\ket{1} \xrightarrow  {T_{13}}  \ket{3} \xrightarrow  {T_{23}} \ket{2} \xrightarrow  {\Omega_{\text{c}}} \ket{3} \xrightarrow {\omega} \ket{1}$, the overall process is as follows: At first, the photon is absorbed from $T_{13}$ reservoir and electron from ground state $\ket{1}$ excites to the excited state $\ket{3}$, as a result, a photon is produced on the $T_{23}$ reservoir, therefore the population of state $\ket{2}$ is increased by one unit. By absorbing a photon from the coupling laser, the atom is excited again from the meta state $\ket{2}$ to the upper level $\ket{3}$. In the final step, a photon having $\omega$ is emitted on the $\ket{3}\rightarrow \ket{1}$ transition, which represents the output of the QHE.
%%%%%%
 %%%%%%%
%%%%%

Hamiltonian for the given system under the rotating wave and dipole approximation is given by
%%%%
\begin{eqnarray}
\label{ham}
H&=&\hbar\Delta\omega_{31}\ket{3}\bra{3}+\hbar\Delta     \omega_{21} \ket{2}\bra{2}- 
      \frac{\hbar}{2}(\text{g}\ket{3}\bra{1}  \nonumber \\ && + \Omega_{\text{c}}\ket{3} \bra{2}+\text{H.c.}),
\end{eqnarray}
%%%
where $\Delta \omega_{31}=\omega_{31}-\omega$ and $\Delta \omega_{21}=\omega_{21}+\omega_{\text{c}}-\omega$ are the detunings of the radiating and control fields, respectively and $\omega_{31} (\omega_{21})$ is the transition frequency between the levels $\ket{3}$ and $\ket{1}$ ($\ket{2}$ and $\ket{1}$). The parameter $\text{g}$ is the coupling coefficient between the emitted field and atomic transition $\ket{3} \rightarrow \ket{1}$. We follow the density matrix approach as all the decay processes are radiative and apply the von Neumann equation to get the equation of motions for density matrix elements.
\begin{equation}
 \dot{\rho}=-\frac{\dot{\iota}}{\hbar}[H, \rho]+ \mathcal L(\rho), 
\end{equation}
where $\mathcal L(\rho)$ is the Liouvillean operator and defined as
\begin{eqnarray}
\mathcal L(\rho)& =& R_{13}(\sigma_{13}\rho\sigma_{31} +\sigma_{31}\rho\sigma_{13}-(\sigma_{33}-\sigma_{11})\rho) \nonumber\\ && + R_{23}(\sigma_{23}\rho\sigma_{32} +\sigma_{32}\rho\sigma_{23}-(\sigma_{33}-\sigma_{22})\rho) \nonumber\\ && +\Gamma_{31}(\sigma_{13}\rho\sigma_{31}-\sigma_{33}\rho) \\ && +\Gamma_{32}(\sigma_{23}\rho\sigma_{32}-\sigma_{33}\rho).
\end{eqnarray}
Where $\sigma_{ij}=\ket{i}\bra{j} (i.j=1,2,3)$,  $\Gamma_{31}$ and $\Gamma_{32}$ are the spontaneous emission rates from the excited state $\ket{3}$ to lower states $\ket{1}$ and $\ket{2}$, respectively. The pumping rates $R_{13} $ and $R_{23}$ can be expressed as
\begin{equation}
    R_{23}=\Gamma_{32} n_{23}=\Gamma_{32}\{\exp{[\hbar \omega_{23}/k_{b} T_{23}}]-1\}^{-1},
\end{equation}
\begin{equation}
     R_{13}=\Gamma_{31} n_{13}=\Gamma_{31}\{\exp{[\hbar \omega_{13}/k_{b} T_{13}}]-1\}^{-1},
\end{equation}
where $T_{13}$ and $T_{23}$ are the ambient temperatures of the thermal radiation fields. The $n_{13}$ and $n_{23}$ are the average number of photons per mode.

The equations of motion for density matrix elements can be written as
\begin{equation}
\label{rate1}
   \dot\rho_{11} = \Gamma_{31} \rho_{33} +\frac{1}{2} \dot{\iota} \text{g} (\rho_{31}-\rho_{13}) + R_{13} (\rho_{33}-\rho_{11}),
\end{equation}
\begin{equation}
 \label{rate2}
       \dot\rho_{22} = \Gamma_{32} \rho_{33}+ \frac{1}{2} \dot{\iota} \Omega_{\text{c}} (\rho_{32}-\rho_{23}) + R_{23} (\rho_{33}-\rho_{22}),
         \end{equation}
\begin{eqnarray}
       \dot\rho_{33}& =& -(\Gamma_{31} +\Gamma_{32}) \rho_{33}+\frac{1}{2} \dot{\iota} \text{g} (\rho_{13}-\rho_{31})+
 \frac{1}{2} \dot{\iota} \Omega_{\text{c}}(\rho_{23}-\rho_{32}) \nonumber\\ &&- R_{23} (\rho_{33}-\rho_{22}) -R_{13}(\rho_{33}-\rho_{11}),
    \end{eqnarray}
    \begin{eqnarray}
        \label{rho12}
 \dot\rho_{12}&=&-(\frac{1}{2} \gamma_{21}-\dot{\iota} \Delta \omega_{21}) \rho_{12}-\frac{1}{2} \dot{\iota} \Omega_{\text{c}} \rho_{13}+\frac{1}{2} \dot{\iota} \text{g} \rho_{32} \nonumber\\ && +\gamma_{s} \rho_{33},
    \end{eqnarray}
\begin{eqnarray}
      \dot\rho_{23}&=&-(\frac{1}{2} \gamma_{32}-\dot{\iota} (\Delta \omega_{31}-\Delta \omega_{21})) \rho_{23}+ \frac{1}{2} \dot{\iota} \Omega_{\text{c}} (\rho_{33}-\rho_{22}) \nonumber\\ && - \frac{1}{2} \dot{\iota} \text{g} \rho_{21},
    \end{eqnarray}
\begin{eqnarray}
\label{rate13}
     \dot\rho_{13}&=&-(\frac{1}{2} \gamma_{31}-\dot{\iota} \Delta \omega_{31}) \rho_{13}-\frac{1}{2} \dot{\iota} \text{g} (\rho_{33}-\rho_{11}) \nonumber\\ && -\frac{1}{2} \dot{\iota} \Omega_{\text{c}} \rho_{12},
\end{eqnarray}
The dephasing rates $\gamma_{21}$, $\gamma_{31}$ and $\gamma_{32}$ are related to decays $\Gamma_{32}$ and $\Gamma_{31}$. For each transition, the dephasing rate is defined as

\begin{equation}
\begin{split}
    \gamma_{21}= R_{13} + R_{23}, \\  \gamma_{31}=\Gamma_{31}+\Gamma_{32}+R_{23}+2 R_{13},\\ \gamma_{32}=\Gamma_{31}+\Gamma_{32}+2R_{23}+R_{13}.\\
\end{split}
\end{equation}
Here we consider SGC between two ground states $\ket{1}$ and $\ket{2}$ \cite{javanainen1992effect, berman2005spontaneously,shui2015efficient}. The factor $\gamma_s$ in equation (\ref{rho12}) represents the coefficient of SGC. The SGC coherence appears due to the cross-coupling between spontaneous emission paths $\ket{3}$ $\rightarrow$ $\ket{1}$ and $\ket{3}$ $\rightarrow$ $\ket{2}$ for non-orthogonal dipole moments. The $\gamma_s$ is defined as \cite{menon1998effects}
\begin{equation}
\label{sgc}
\gamma_{s}=\frac{|\Vec{d}_{31}.\Vec{d}_{13}|}{|\Vec{d}_{32}| |\Vec{d}_{31}|}\sqrt{\Gamma_{31}\Gamma_{32}}/2=\text{p} \sqrt{\Gamma_{31}\Gamma_{32}}/2, 
\end{equation}
 here $|\Vec{d}_{31}| $ and $|\Vec{d}_{32}|$ are the dipole moments and p= $\cos{\theta}$, where $\theta $ is the angle between these dipole moments, see Fig. \ref{fig:model}(b). SGC effect disappears when these dipole moments are orthogonal \cite{javanainen1992effect,niu2006enhancing}. In our proposed system, we vary the values of the parameter p, which is equal to the $\cos{\theta}$. By doing so, we investigate the impact of different values of p on the dynamics of the QHE. 
 
 In the absence of any reflection or scattering to other modes, the spectral brightness as a function of $z$ an be written as \cite{harris2016electromagnetically}
\begin{equation} \label{ode1}
    \frac{dB(\omega,z)}{dz}+N[\sigma_{A} \rho_{11}-\sigma_{E}(\rho_{22} + \rho_{33})]B(\omega,z)= \sigma_{E}(\rho_{22} + \rho_{33}).
 \end{equation}
Here, $N$ is a number density, $\sigma_{A}$ and $\sigma_{E}$ are absorption and emission cross-sections, respectively. The brightness is zero at $z=0$, and becomes maximum $B_{\text{black}}(\omega)$ at a certain value of $z$, where all relevant spectral components are fully absorbed. The brightness for each spectral component can be determined by evaluating the steady state solution of Eq. (\ref{ode1}) ($\frac{dB(\omega,z)}{dz}=0$), as expressed by the following equation
\begin{equation}
    B_{\text{black}}(\omega)=\frac{\Lambda \sigma_{E}}{\sigma_{A}-\Lambda \sigma_{E}},
    \label{B}
\end{equation}
here, $\Lambda$ is defined as the "ratio of the number of atoms in the upper state to those in the ground state". In the weak field limit, i.e., $\text{g}<\Omega_{c}$, $\Lambda$ can be defined as $\Lambda={\rho_{22}^{(0)}+\rho_{33}^{(0)}}/{\rho_{11}^{(0)}}$, where $\rho_{ii}^{(0)}(i=1,2,3)$ is the first solution given in Appendix \ref{appendixA}. 
The absorption $\sigma_{A}$ and emission $\sigma_{E}$ cross sections can be obtained by solving the density matrix equations (Eqs. (\ref{rate1}-\ref{rate13}) under steady state condition. The $\sigma_{A}$ and $\sigma_{E}$ are related to the imaginary part of the first order solution of $\rho_{13}^{(1)}$ \cite{imamoglu1991lasers}, i.e.,
\begin{equation}
\label{stand}
    Im (\text{g}^* \rho_{13}^{(1)})=-\sigma_{A} 
    \rho_{11}^{(0)}+\sigma_{E}(\rho_{22}^{(0)}+\rho_{33}^{(0)}),
\end{equation}
where the expression for $\rho^{(1)}_{13}$ is given in appendix \ref{appendixB}.
%%%%
\begin{figure}[t]
    \centering
    \includegraphics[width=8.6cm]{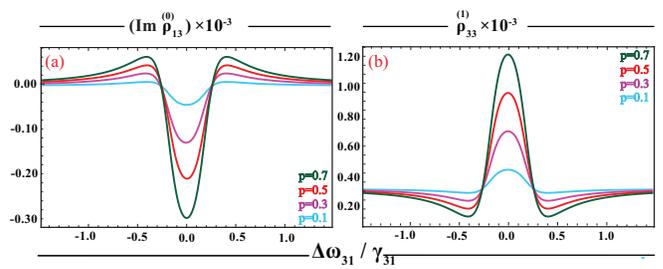}
    \caption{(a) Gain profile Im$[\rho^{(0)}_{13}]$ and (b) excited-state population $\rho^{(1)}_{33}$ is plotted against detuning $\Delta \omega_{31}$ for different values of p. The remaining parameters are $\Gamma_{31}=10^7, \Gamma_{32}=6 \times10^7,  \Omega_{\text{c}}=5\times 10^7, \omega_{13}= 4\times 10^{15},  \omega_{12}=10^{15}, \Delta \omega_{21}=0$, $T_{23}=5778K$, $T_{13}=3778K$ and $\Lambda=0.019$.}
    \label{fig:coherence}
\end{figure}
%%%

First, we consider the role of SGC on the coherence between the ground state ($|1\rangle$) and the excited state ($|3\rangle$) and the population of the upper level $|3\rangle$. In Fig. \ref{fig:coherence}, we plot Im$[\rho^{(0)}_{13}]$ and $\rho^{(1)}_{33}$ against detuning for different values of p. The remaining parameters are $\Gamma_{31}=10^7, \Gamma_{32}=6 \times10^7,  \Omega_{\text{c}}=5\times 10^7, \omega_{13}= 4\times 10^{15},  \omega_{12}=10^{15}, \Delta \omega_{21}=0$, 
$\Lambda=0.019$ and $T_{23}=5778K$, $T_{13}=3778K$. The Fig. \ref{fig:coherence}(a) illustrates the gain (Im$[\rho^{(0)}_{13}]<0$) in the system prior to the generation of the output field and this gain increases with the increase in SGC. The graph demonstrates a continuous enhancement in gain at resonance $\Delta\omega_{31}=0$ as the value of the parameter p increases from $0.1$ to $0.7$. The gain is induced in the system without inversion as most of the atoms are in the ground state and with only a few atoms found in the excited state $|3\rangle$ as shown in Fig. \ref{fig:coherence}(b). In fact, even for p=$0.7$, the population of the excited state $\rho^{(1)}_{33}$ is low as $1.2 \times 10^{-3}$, as illustrated with green curve in Fig. \ref{fig:coherence}(b). The gain that arises due to SGC is crucial for the operation of the EIT-based QHE. The SGC plays a significant role in enhancing the emission and spectral brightness of the system by generating coherence between lower levels.
%%%%%%%
\section{Absorption and Emission cross-sections}
%%%%%%%%
In this section, we analyze the effect of the SGC on the absorption and emission cross-section. The SGC coherence arises due spontaneous decay rates $\Gamma_{31}$ and $\Gamma_{32}$ from the excited state $\ket{3}$ to lower states $\ket{1}$ and $\ket{2}$, see Eq. (\ref{sgc}). It only affects the spontaneous emission of the field. The absorption cross-section $\sigma_{A}$ depends on the ground state population $\sigma_{A} \rho_{11}^{(0)}$ while emission-cross $\sigma_{E}$ is related to the excited state population as given in Eq. (\ref{stand}). The SGC has no effect on the absorption of the medium and most of the atoms remain in the ground state. We obtain the same result for absorption cross-section as given in \cite{harris2016electromagnetically}. However, SGC amplifies the generated spontaneous radiation and significantly changes the emission cross-section. The expression for emission cross-section is lengthy and complex, making it unsuitable to present here. It is provided in Eq. \ref{sgc eq} of Appendix \ref{appendixB}. However, we can write $\sigma_{E}$ as 
\begin{equation}
    \sigma_{E}/\sigma_{0}= \sigma_{EIT}+\sigma_{SGC},
\end{equation}
  where,
   $\sigma_{0}=(2 \text{g} |d_{13}|^2)/(\epsilon c \hbar \gamma_{13})$, $\sigma_{EIT}$ is same as given in \cite{harris2016electromagnetically}. The $\sigma_{SGC}$ (see Eq. \ref{sgc eq}) shows a quadratic enhancement with respect to SGC coefficient $\gamma_{s}$. This suggests a non-linear dependence between the SGC coefficient $\gamma_{s}$ and $\sigma_{SGC}$ which shows that the $\gamma_{s}$ has a greater impact on $\sigma_{SGC}$ at higher values.

\begin{figure}[t]
    \centering     \includegraphics[width=3.4in]{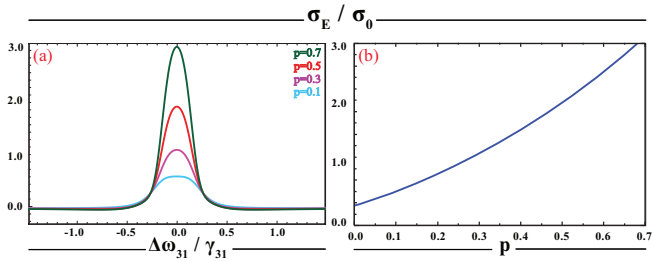}
    \caption{Emission cross section is plotted against the detuning $\Delta \omega_{31}$ for p=0.1 (blue curve), p=0.3 (purple curve), p=0.5 (red curve) and  p=0.7 (green curve). (b) Normalized emission cross-section vs p (spontaneously generated factor). Other parameters are the same as given in Fig. \ref{fig:coherence}. }
    \label{fig:em sgc}
\end{figure}

In Fig. \ref{fig:em sgc}, we plot the emission cross section against the detuning for four different values of p, i.e, p=$0.1$, p=$0.3$, p=$0.5$ and p=$0.7$. The remaining parameters are the same as given in Fig. \ref{fig:coherence}.
%%%%%
 %%%%%%
  Our results demonstrate a significant enhancement in the emission cross-section as the value of p is varied as shown in Fig \ref{fig:em sgc}. At a value of p=0.1, the emission cross-section exhibits a value of $0.56\sigma_{0}$. Upon increasing the value of p from 0.1, there is a corresponding increase in the emission cross-section. For example, at p=0.3 the emission cross-section reaches $\sigma_{0}$ depicted by a purple line. Further increases in p to 0.5 result in a further increase in the emission profile, reaching $1.8\sigma_{0}$ shown by the red curve. Finally, when p=0.7, the peak value of the emission cross-section approaches $3\sigma_{0}$. The fact that the emission cross-section can be made greater than 1 through tuning p indicates the presence of a gain in the system.
For further validation, the emission cross-section as a function of p is plotted \ref{fig:em sgc} (b).  An analysis of the normalized emission cross-section in relation to the parameter p demonstrates a gradual rise in the emission spectrum from its initial value as p increases. The emission spectrum attains its peak value of $3\sigma_{0}$ at p=0.7, demonstrating a strong correlation between the emission cross-section and p, which is very helpful for enhancing the brightness (output) of QHE.
%%%%
\begin{figure}[t]
    \centering
    \includegraphics[width=3.4in]{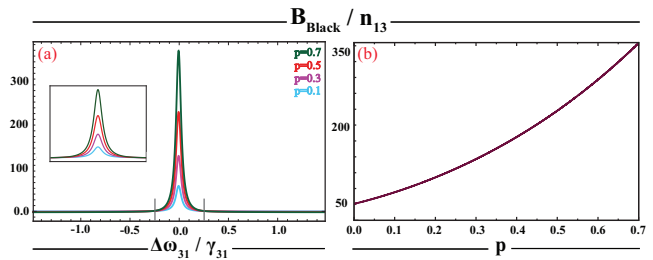}
    \caption{(a) Normalized brightness is plotted against detuning for different values of p such as p=0.1 (blue curve), p=0.3 (purple curve), p=0.5 (red curve), and p=0.7 (green curve). The interpretation of the graphs shows that for increasing values of p, we get enhancement in the brightness profile. The inset is the zoomed-in region from $\Delta \omega=-0.5\gamma_{31}$ to $\Delta \omega=0.5\gamma_{31}$. (b) Normalized spectral brightness is plotted as a function of p. Other parameters are the same as given in Fig. \ref{fig:coherence}. }
    \label{fig:bright}
\end{figure} 
%%%
%%%%%%%%
 \section{Spectral Brightness}
 Now, we investigate the spectral brightness which is related to the output power of the QHE.
The results are depicted in Fig. \ref{fig:bright}(a), where the brightness spectrum is plotted as a function of detuning $\Delta\omega_{31}$ for four distinct values of the parameter p: p=0.1, p=0.3, p=0.5, and p=0.7. Fig. \ref{fig:bright} shows that an increase in the value of p leads to a sharp peak in the brightness spectrum. Additionally, the intensity of the spectral brightness experiences a significant increase as p increases. The spectral brightness reaches a maximum value of $370 n_{13}$ when p=0.7, as indicated in Fig. \ref{fig:bright} (a).
Moreover, a graph is plotted to investigate the correlation between normalized spectral brightness and the parameter p, see Fig. \ref{fig:bright} (b). The results indicate an increase in spectral brightness with a corresponding increase in the value of p. These results are significant as they demonstrate the potential for a gain-assisted QHE to produce a large output through the implementation of SGC. They also provide evidence that by inducing coherence, the performance of a QHE can be enhanced to a certain extent.
%%%%%
\begin{figure}[t]
 \includegraphics[width=3.4in]{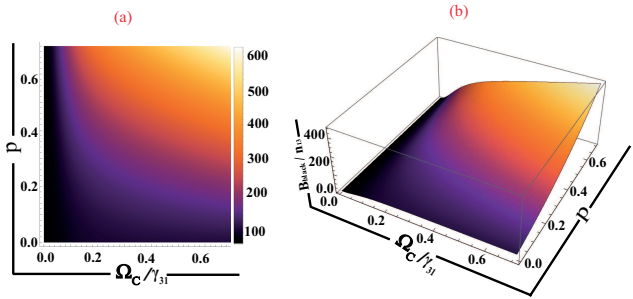}
    \caption{(a) A density plot of spectral brightness is plotted as a function of two variables, p, and $\Omega_{\text{c}}$. (b) A 3D plot of spectral brightness is generated against the same variables, p, and $\Omega_{\text{c}}$, while the remaining parameters are same as given in Fig. \ref{fig:coherence}. }
    \label{fig:density}
\end{figure}
%%%%%
%%%%%

Next In Fig. \ref{fig:density}, we analyze the dependence of the spectral brightness on the parameters p and control field $\Omega_{\text{c}}$. A three-dimensional density graph and a three-dimensional plot are generated to visualize this relationship. The results clearly demonstrate that the spectral brightness increases in tandem with the increase of both p and $\Omega_{\text{c}}$. The graphical representation provides a comprehensive understanding of the dependence of spectral brightness on the two parameters.
%%%
The results from Fig. \ref{fig:density} (a) indicate that the atomic system exhibits its maximum spectral brightness of $370 n_{13}$ when we set p=0.7 and the control field $\Omega_{\text{c}}=0.7\gamma_{31}$. These results were further supported by the 3D plot in Fig. \ref{fig:density} (b), which demonstrates that the highest spectral brightness is achieved at the same values of $\Omega_{\text{c}}=0.7\gamma_{31}$ and p=0.7. Moreover, these two graphs support the fact that the brightness and emission cross-section at p = 0, is consistent with a typical Harris heat engine.

 The efficiency of our proposed gain-assisted QHE is the same as EIT-based QHE described in Ref. \cite{harris2016electromagnetically}. However, due to the effect of the SGC, the brightness, which is related to the power of the QHE, is greatly enhanced. These particular QHEs can be extremely important in many different ways. They might be utilized to move energy in nano-scale devices or to cool down quantum systems also they can be use in other fields such as quantum sensing, quantum communication, quantum information processing, and quantum metrology.
In our situation, the typical Harris heat engine's power can be boosted up to a certain point, and this power can be used to harness work for potential applications.
%%%%%
\section{Conclusion}
 We have proposed a scheme to realize a gain-assisted QHE based on EIT. We utilized the SGC which is generated between two lower levels due to the cross-coupling between the two spontaneous pathways from an excited state. The emission cross-section and brightness strongly depend on the SGC. The SGC induced gain without inversion in the system and as a result, both the emission-cross section and brightness are enhanced. In summary, the output power in our proposed EIT-Based QHE can be increased with the aid of SGC.
\begin{widetext}
\appendix 
\section{Zero and First-order Solutions} \label{appendixA}
To calculate the absorption and emission cross-sections, the density matrix equations are solved in a perturbative manner. In the linear regime, we take $\rho_{ij}= \rho^{(0)}_{ij}+\lambda \rho^{(1)}_{ij} (i,j=1,2,3)$, where $\lambda$ is a small number representing the interaction order. We assume that radiating field is weak and choose $\text{g}=\lambda\text{g}$ and consider the resonant interaction of the coupling field with atom $\Delta\omega_{21}=0$. The steady-state solution $\dot\rho_{ij}=0$ of the density matrix elements (Eqs (\ref{rate1})-(\ref{rate13})) to all order of $\Omega_{c}$ and zero-order of $\text{g}$ are given by
\begin{align}
    \rho^{(0)}_{11} =\frac{(R_{13} + \Gamma_{31}) (R_{23} \gamma_{32} +\Omega_{\text{c}}^2)}{
 R_{23} \Gamma_{31} \gamma_{32} + 
  R_{13} \Gamma_{32} (3 R_{23} + \Gamma_{32}) + (3 R_{13} + 
\Gamma_{31}) \Omega_{\text{c}}^2},
\end{align}
\begin{align}
   \rho^{(0)}_{22} = \frac{
 R_{13} (\Gamma_{32} (R_{23} + \Gamma_{32}) + \Omega_{\text{c}}^2)}{(
 R_{23} \Gamma_{31} \Gamma_{32} + 
  R_{13} \Gamma_{32} (3 R_{23} + \Gamma_{32}) + (3 R_{13} + \
\Gamma_{31}) \Omega{23}^2)},
\end{align}

\begin{align}
   \rho^{(0)}_{33} = \frac{R_{13} (R_{23} \gamma_{32} + \Omega_{\text{c}}^2)}{
 R_{23} \Gamma_{31} \gamma_{32} + 
  R_{13} \gamma_{32} (3 R_{23} + \Gamma_{32}) + (3 R_{13} + \
\Gamma_{31}) \Omega_{\text{c}}^2},
\end{align}
\begin{align}
   \rho^{(0)}_{23}= \frac{-\dot \iota R13 \Gamma_{32} \Omega_{\text{c}}}{
  R_{23} \Gamma_{31} \gamma_{32} + 
   R_{13} \Gamma_{32} (3 R_{23} + \Gamma_{32}) + (3 R_{13} + \
\Gamma_{31}) \Omega_{\text{c}}^2},
\end{align}
\begin{equation}
      \rho^{(0)}_{12}=\frac{
 2 R_{13} \gamma_{s} (\gamma_{31} - 
    2 \dot \iota \Delta \omega_{31}) (R_{23} \gamma_{32} + \
\Omega_{\text{c}}^2)}{((\gamma_{21} - 
       2 \dot \iota \Delta\omega_{31}) (\gamma_{31} - 
       2 \dot \iota \Delta\omega_{31}) + \Omega_{\text{c}}^2) (R_{23} 
\Gamma_{31} \gamma_{32} + 
    R_{13} \gamma_{32} (3 R_{23} + \Gamma_{32}) + (3 R_{13} + \
\Gamma_{31}) \Omega_{\text{c}}^2)},
\end{equation}
\begin{equation}
    \rho^{(0)}_{13}= \frac{-
  2 \dot \iota 
    R_{13} \gamma_{s} \Omega_{\text{c}} (R_{23} \gamma_{32} + \
\Omega_{\text{c}}^2)}{((\gamma_{21} - 
        2 \dot \iota \Delta\omega_{31}) (\gamma_{31} - 
        2 \dot \iota \Delta \omega_{31}) + \Omega_{\text{c}}^2) (R23 \
\Gamma_{31} \gamma_{32} + 
     R_{13} \gamma_{32} (3 R_{23} + \Gamma_{32}) + (3 R_{13} + \
\Gamma_{31}) \Omega_{\text{c}}^2)}.
\end{equation}
Express $\rho^{(0)}_{13}$ and $\rho^{(0)}_{12}$ in terms of $\rho^{(0)}_{33}$ to facilitate the calculation of emission cross-section and spectral brightness.

\begin{align}
 \label{rho122} \rho^{(0)}_{12}=\frac{2 \gamma_{s}  \rho^{(0)}_{33} (\gamma_{31}-2 \dot \iota \Delta \omega_{31}) }{[(\gamma_{31}-2 \dot \iota \Delta \omega_{31})(\gamma_{21}-2 \dot \iota \Delta \omega_{31})+\Omega_{\text{c}}] },  
\end{align}
\begin{align}
  \label{rho133} \rho^{(0)}_{13}=\frac{-2 \dot \iota \gamma_{s} \rho^{(0)}_{33} \Omega_{\text{c}}} {[(\gamma_{31}-2 \dot \iota \Delta \omega_{31})(\gamma_{21}-2 \dot  \iota \Delta \omega_{31})+\Omega_{\text{c}}] },  
\end{align}
%%%%%
The steady-state solution for $\rho_{33}$ and $\rho_{13}$ to all order of $\Omega_{c}$ and first order of $\text{g}$ are obtained as
\begin{align}
  \label{rho333}  \rho^{(1)}_{33}=  \rho^{(0)}_{33} +\frac{(\rho^{(0)}_{21} + \rho^{(0)}_{12}
     ) R_{13} \Omega_{\text{c}} \text{g} + 
  \dot \iota (\rho^{(0)}_{13}-\rho^{(0)}_{31} )(R_{23} \gamma_{32} + \Omega_{\text{c}}^2) \text{g}}{
 2 \gamma_{32} (R_{23} \Gamma_{31} + 
     R_{13} (3 R_{23} + \Gamma_{32})) + 
  2 (3 R_{13} + \Gamma_{31}) \Omega_{\text{c}}^2},
\end{align}
\begin{align}
  \label{rho131}   \rho^{(1)}_{13}= \frac{\dot \iota \text{g} \Omega_{\text{c}}^2 (\rho^{(0)}_{22}-\rho^{(0)}_{33}) +2 \dot \iota \Omega_{\text{c}} \gamma_{s} \gamma_{32} \rho^{(1)}_{33} +  \dot \iota \gamma_{32}(\gamma_{21}-2 \dot \iota \Delta \omega_{31}) \text{g}(\rho^{(0)}_{33}-\rho^{(0)}_{11})}{[(\gamma_{31}-2 \dot \iota \Delta \omega_{31})(\gamma_{21}-2 \dot \iota \Delta \omega_{31})+\Omega_{\text{c}}^2]}.
\end{align}
\section{Expression for Emission Cross-Section } \label{appendixB}
The detailed and step-by-step calculation for absorption and emission cross-sections without SGC can be found in Ref. \cite{imamoglu1991lasers}. Here, these cross-sections can be obtained from the imaginary part of $\rho^{(1)}_{13}$, see Eq. (\ref{stand}). By substituting Eqs. (\ref{rho122})-(\ref{rho333}) in Eq. (\ref{rho131}) and and performing algebraic manipulations, one can express Im$[\rho^{(1)}_{13}]$ as a function of the diagonal terms, $\rho^{(0)}_{11}$, $\rho^{(0)}_{22}$ and $\rho^{(0)}_{33}$. The coefficient of $\rho^{(0)}_{11}$ gives the absorption cross-section which is the same as given in \cite{harris2016electromagnetically}. The coefficient associated with $(\rho_{22}^{(0)}+\rho_{33}^{(0)})$ corresponds to the emission cross-section that consists of two parts. The first part represents the emission of EIT-based QHE as described in the work of Harris \cite{harris2016electromagnetically}. The second part represents the emission cross-section incorporating the effect of SGC. The combination of these two emission cross-sections provides the overall expression for gain-assisted QHE based on EIT.
\begin{equation}
    \sigma_{E}/\sigma_{0}= \sigma_{EIT}+\sigma_{SGC},
\end{equation}
  where,
   $\sigma_{0}=(2 \text{g} |d_{13}|^2)/(\epsilon c \hbar \gamma_{13})$ and
  \begin{align}
      \sigma_{SGC}=
\frac{\gamma_{31}}{\text{g}} \left[  \gamma_{s}^{2} \left( 2 \Omega_{\text{c}} \left( \left( R_{23} \gamma_{32} + \Omega_{\text{c}}^{2} \right) \left( \text{g} \Omega_{\text{c}} \mathcal{A} + \text{g} \Omega_{\text{c}}\left(\mathcal {B}+\mathcal{C} \right) \right) \right) \right)  +\gamma_{s} \mathcal{D}\right],
      \label{sgc eq}
  \end{align}
 where
$$\mathcal A = (R_{13} \gamma_{31} + R_{23} \gamma_{32} + \Omega_{\text{c}}^2) ((\gamma_{21}^2 + 4 \Delta \omega^{2}_{31}) (\gamma_{31}^2 + 4 \Delta \omega^{2}_{31}) + 2 (\gamma_{21} \gamma_{31} - 4 \Delta \omega^{2}_{31}) \Omega_{\text{c}}^2 + \Omega_{\text{c}}^4), $$

$$\mathcal B = (R_{23} \gamma_{32} + \Omega_{\text{c}}^2) (\gamma_{21} \gamma_{31} - 2 (\gamma_{21} + \gamma_{31}) \Delta \omega_{31} - 4 \Delta \omega^{2}_{31} + \Omega_{\text{c}}^2) (\gamma_{21} \gamma_{31} + 2 (\gamma_{21} + \gamma_{31}) \Delta \omega_{31} - 4 \Delta \omega^{2}_{31} + \Omega_{\text{c}}^2), $$
\begin{eqnarray*}
\mathcal{C} & =& R_{13} (\gamma_{21}^2 \gamma_{31}^3 - 12 \gamma_{21}^2 \gamma_{31} \Delta \omega^{2}_{31} - 24 \gamma_{21} \gamma_{31}^2 \Delta \omega^{2}_{31} - 4 \gamma_{31}^3 \Delta \omega^{2}_{31} + 32 \Gamma_{21} \Delta \Omega^{4}_{31} + 48 \Gamma_{31} \Delta \omega^{4}_{31} \nonumber\\ && + 2 (\gamma_{21} \gamma_{31}^2 - 4 (\gamma_{21} + 2 \gamma_{31}) \Delta \omega^{2}_{31}) \Omega_{\text{c}}^2 + \gamma_{31} \Omega_{\text{c}}^4),
\end{eqnarray*}
$$ \mathcal{D} = \frac{2 \Omega_{\text{c}} (R_{23} \gamma_{32} + \Omega_{\text{c}}^2) (\gamma_{21} \gamma_{31} - 4 \Delta \omega^{2}_{31} + \Omega_{\text{c}}^2)}{(2 R_{23} \gamma_{32} + \gamma_{32} \gamma_{32} + 2 \Omega_{\text{c}}^2) ((\gamma_{21}^2 + 4 \Delta \omega^{2}_{31}) (\gamma_{31}^2 + 4 \Delta \omega^{2}_{31}) + 2 (\gamma_{21} \gamma_{31} - 4 \Delta \omega^{2}_{31}) \Omega_{\text{c}}^2 + \Omega_{\text{c}}^4)}. $$
The expression for brightness is complex and is not provided in this paper, but it can be determined by utilizing Eq. \ref{B}.

\end{widetext}
\section*{Acknowledgments}
The research presented in this work has been conducted under the support of the Higher Education Commission (HEC) of Pakistan's National Research Program for Universities (NRPU) funded project entitled "Realization of Optical Devices using Quantum Dots Rydberg Atoms and 2D Graphene structure" with grant agreement No. 10692/Balochistan/NRPU/R and D/HEC/2017.
\bibliography{bibliography}
\bibliographystyle{apsrev4-1}
\end{document}